\documentclass[fleqn,twoside]{article}
\usepackage{espcrc2}
\usepackage{epsfig}

\usepackage{graphicx}
\newcommand{\AmS}{{\protect\the\textfont2
  A\kern-.1667em\lower.5ex\hbox{M}\kern-.125emS}}

\title{Leptogenesis with Majorana neutrinos}

\author{E.A.\ Paschos\address{Institut f\"{u}r Physik, Universit\"{a}t Dortmund\\
D-44221 Dortmund, Germany}\thanks{Invited Talk presented at the
Workshop on Low Energy Neutrino--Nucleus Interactions, Dec. 13 -- 16,
2001, KEK, Tsukuba, Japan.}}
       
\begin{document}
\begin{abstract}
I review the origin of the lepton asymmetry which is
converted to a baryon excess at the electroweak scale.
This scenario becomes more attractive if we can relate it 
to other physical phenomena. For this reason I elaborate
on the conditions of the early universe which lead to
a sizable lepton asymmetry.  Then I describe methods and
models which relate the low energy parameters of neutrinos
to the high energy (cosmological) CP--violation and to
neutrinoless double $\beta$--decay.
\vspace{1pc}
\end{abstract}

\maketitle

\section{What is known about Neutrinos?}

Extensive studies with neutrinos have  established many 
of their properties.
Neutrino--induced processes are either 
leptonic or semileptonic. In the semileptonic processes
there are also hadronic matrix elements whose properties
are known to various degrees of accuracy.  
This was one of the 
main topics of this conference: to report and compare
various calculations at the few percent level.
I must also say that there are many calculations waiting
to be compared with the data (low energy 
$\Delta$--resonance,
nuclear target effects, etc.).
I will cover several reactions in my second talk to
this conference. 

Beyond their couplings, neutrinos have special properties.
\begin{enumerate}
\item[i)] Neutrinos of various generations mix 
with each other, implying that there are physical states
of different masses.
\item[ii)]  It is possible to construct coherent states
of particles and antiparticles, known as Majorana 
neutrinos.
\end{enumerate}

The above properties allow us to write two
kinds of mass terms:
Dirac $M_D\,\bar{\nu}_L\, \nu_R$ and
{\mbox Majorana} masses $M_L\, \bar{\nu}_L^c\, \nu_L$ and 
$M_R\, \bar{\nu}_R^c\, \nu_R$.

The latter allow the states:
$N\propto \nu_R + \nu_R^c$ and
$\nu_M \propto \nu_L + \nu_L^c$
with which we can write the mass matrix
\begin{displaymath}
\left(\bar{\nu}_M\quad\bar{N}\right)
\left( \begin{array}{cc}
   m_L & m_D \\ m_D & M_R  \end{array} \right)
\left( \begin{array}{c}
    \nu_M \\ N \end{array} \right)\, .
\end{displaymath}

The mixing phenomena observed so far are sensitive only to
mixings among
generations and say nothing about the presence of Majorana 
mass terms.  We may now ask if the Majorana nature of 
particles influences other phenomena, like
Leptogenesis, Baryogenesis, Neutrinoless Double Beta 
Decay, etc.

I will describe next how the mixing and the decays
of Majorana-type
neutrinos produces an asymmetry between leptons and antileptons
in the universe.  This phenomenon provides an attractive
scenario for the generation of a lepton asymmetry in the
early universe, which was converted, at a later epoch, 
into a baryon-asymmetry.

\section{Lepton Asymmetry for Heavy Majorana Neutrinos}

We extend the standard model to include a right--handed 
neutrino in each generation \cite{Fukuyana}.  A typical generation is
\begin{equation}
\psi_L = \left( \begin{array}{c}
  \nu_e \\ e^- \end{array}\right)_L\, ,
\quad e_R\, ,\quad (N_e)_R
\end{equation}
We are led to an $SU(2)\times U(1)$ invariant 
Lagrangian
\begin{eqnarray}
{\cal{L}}_F & = & i\bar{\psi}_R\gamma_{\mu}
 (\partial^{\mu}+i\,g'\frac{y}{2}\,B_{\mu})
   \psi_R\nonumber\\
& + & i\bar{\psi}_L\gamma_{\mu}(\partial^{\mu}+i\, g'
   \frac{y}{2}\, B^{\mu}+ i\, g\,\tau^k A^{k,\mu})
     \psi_L\nonumber\\
& + & h_{\alpha\beta}\,\bar{\psi}_L^{\alpha}
  \left( \begin{array}{c}
   \phi^+\\ \phi^0+v \end{array}\right)
   \psi_R^{\beta}\nonumber\\
& & \left[{\rm providing}\,\,{\rm mixing}\,\,{\rm of}\,\,
{\rm the}\,\,{\rm families}\right.\,\,\nonumber\\
& & \left. {\rm sufficient}\,\,
{\rm for}\,\,{\rm oscillations}\right]\nonumber\\
& + & k_{ij}\,\bar{\psi}_R^{c,i}\psi_R^j(H^0+w)\nonumber\\
& & \left[{\rm requires}\,\,{\rm a}\,\,{\rm new}\,\,
{\rm Higgs}\,\,{\rm singlet}\right]
\end{eqnarray}
This theory has new couplings like:
$h_{\alpha_i}\,\bar{\ell}_{L\alpha}\phi N_{Ri}$
and a mass term of Majorana type:
$k_{ij}\, \bar{\psi}_R^{c,i}\psi_R^j w\, .$

We select the Majorana mass matrix to be real and 
diagonal.  The new couplings introduce new diagrams in
decays and in the mass terms.
Decays like

%DIAGRAMS (Fig.1)
\begin{figure}[h]
\centerline{
\begin{minipage}[r]{2.8cm}
\epsfig{file=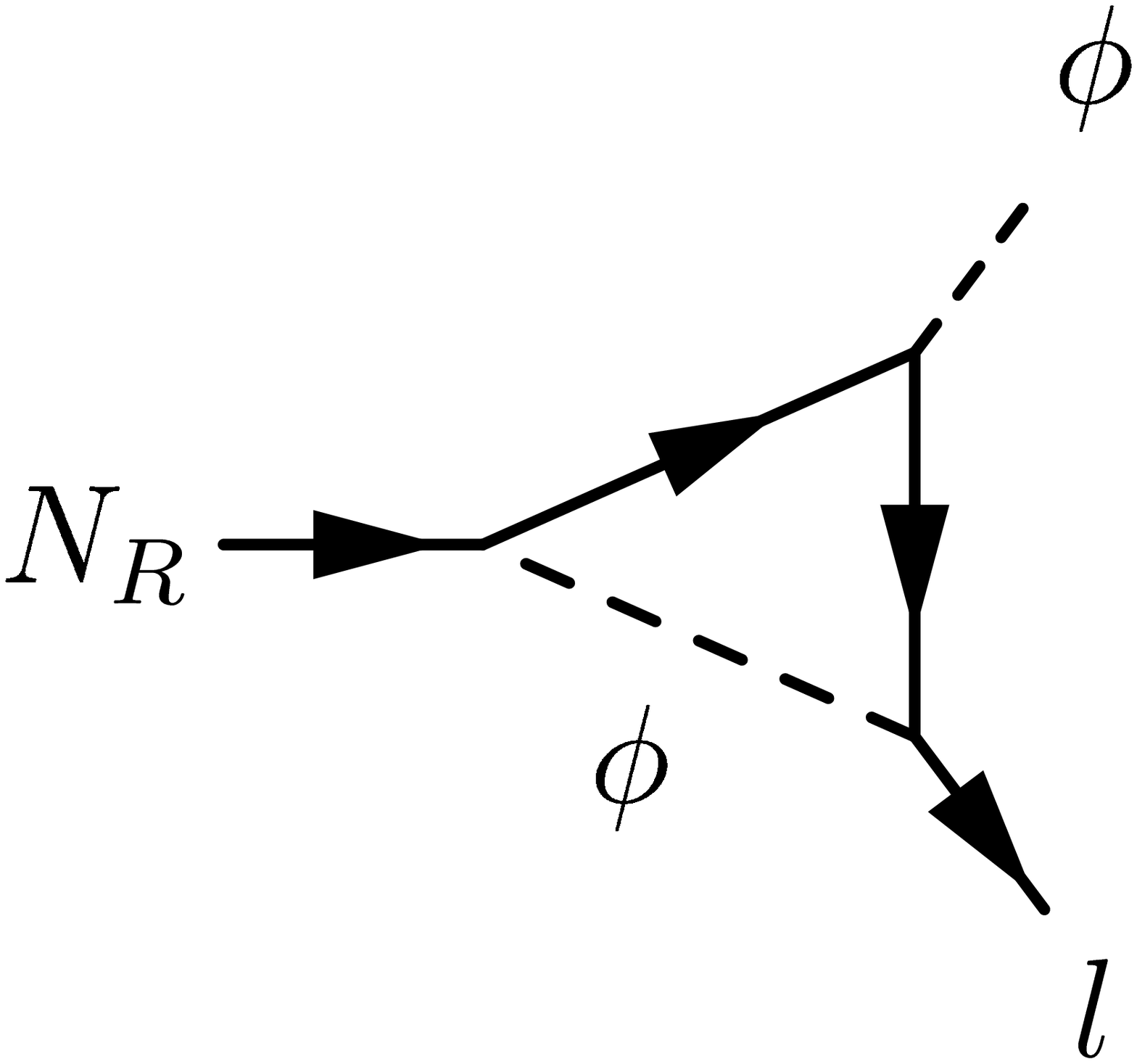,width=2.8cm}
\end{minipage}
\hfill
\begin{minipage}[c]{0.4cm}
+
\end{minipage}
\hfill
\begin{minipage}[l]{2.8cm}
\epsfig{file=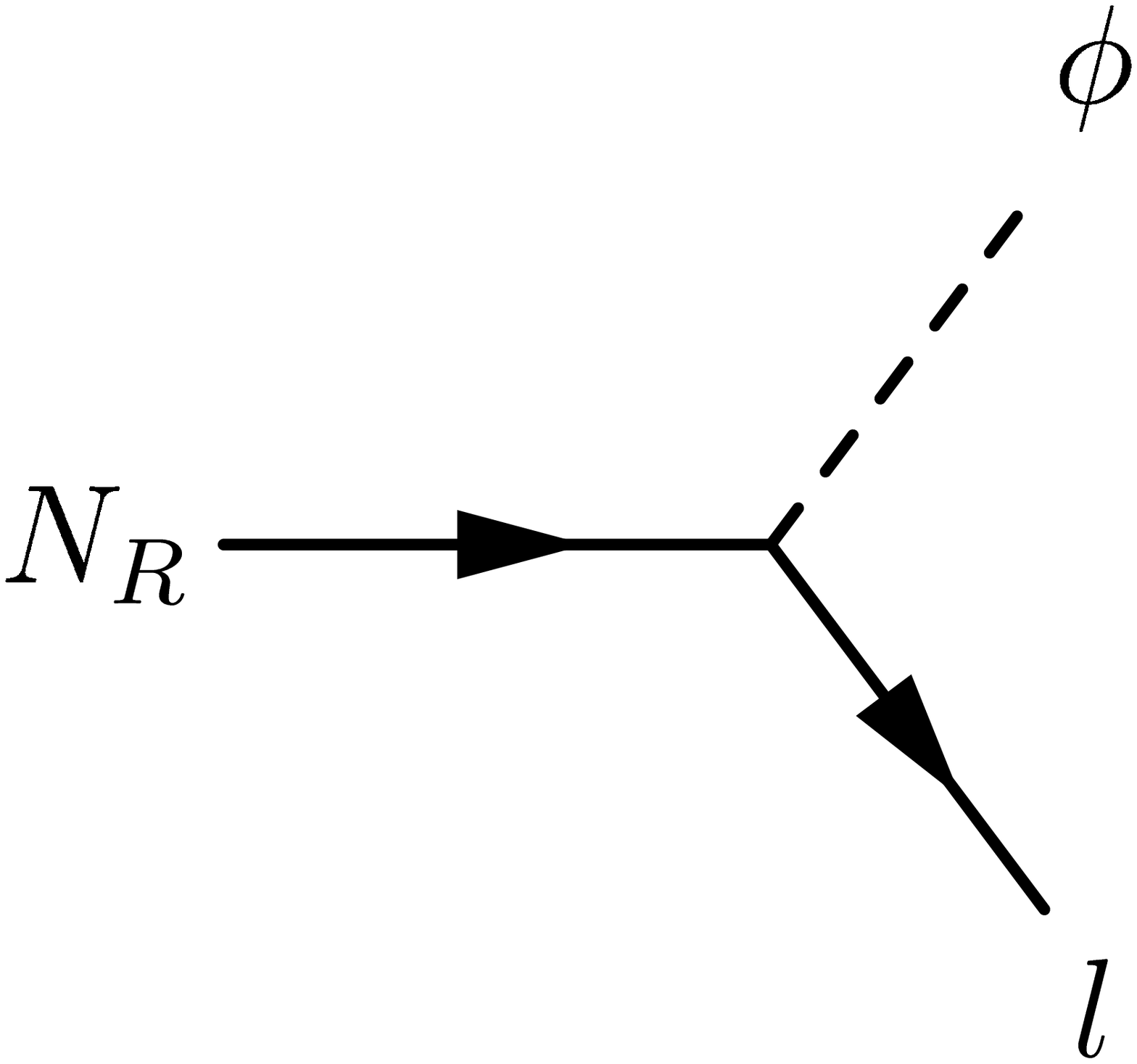,width=2.8cm}
\end{minipage}}
\caption{Born and vertex diagrams}
\end{figure}
\noindent and the mixing of states in the mass matrix

%\DIAGRAM (Fig. 2)
\begin{figure}[h]
\centerline{\epsfig{file=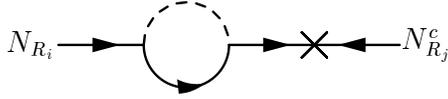,width=6cm}}
\caption{Self--energy}
\end{figure}
\noindent produce an asymmetry
\begin{eqnarray}
\varepsilon & = & \frac{\Gamma(N_{R_i}\to\ell)-
   \Gamma(N_{R_i}\to\ell^c)}
   {\Gamma(N_{R_i}\to\ell)+
    \Gamma(N_{R_i}\to\ell^c)}\\
&  = & \frac{1}{8\pi v^2}
     \frac{1}{(m_D^+ \,m_D)_{11}}
      \sum_{j=2,3} Im\left((m_D^+\, m_D)^2_{1j}\right)f(x)\nonumber
\end{eqnarray} 
with
$f(x) = \sqrt{x}\left\{ \frac{1}{1-x}+1-(1+x)\ln
    \left(\frac{1+x}{x}\right)\right\}$,
$x=\left(\frac{M_j}{M_1}\right)^2$ and $M_1$ the mass
of the lightest Majorana neutrino. The term
$\left(\frac{1}{1-x}\right)$ comes from the mixing of
the states \cite{Flanz95} and the rest from the interference of 
vertex corrections with Born diagrams
\cite{Fukuyana,Luty}. 
The above formula is an approximation for the case 
when the two masses are far apart.  In case they are
close together there is an exact formula, showing 
clearly a resonance phenomenon \cite{Flanz96} 
from the overlap of the two wave functions.
The origin of the asymmetry has also been studied in
field theory \cite{Pilaftsis,Buch} and supersymmetric theories
\cite{Covi}.

Consider a theory with two generations.  The current
states for the neutrinos and antineutrinos are
represented by the column matrix
\begin{displaymath}
\left( \begin{array}{c}
  N_1^c \\ N_1 \\ N_2^c \\ N_2 \end{array}\right)\, .
\end{displaymath}
On this basis higher order corrections produce the
mass matrix
\begin{displaymath}
\left( \begin{array}{cccc}
  0 & M_1+H_{11} & 0 & H_{12}\\
  M_1+H_{11} & 0 & \tilde{H}_{12} & 0\\
  0 & \tilde{H}_{12} & 0 & M_2+H_{22}\\
  H_{12} & 0 & M_2+H_{22} & 0 \end{array}\right)
\end{displaymath}
whose details are given in article \cite{Flanz96}. 
The physical states are now superpositions of particles
and antiparticles
\begin{eqnarray*}
|\psi_1\rangle & = & C_1 \left[a_1 N_1^c+b_1 N_2^c 
      +c_1 N_1 + d_1 N_2\right]\\
|\psi_2\rangle & = & C_2 \left[a_2 N_1^c +b_2 N_2^c 
      +c_2 N_1 + d_2 N_2\right].
\end{eqnarray*}
$a_i, \ldots d_i$'s are constants from the solution of
the eigenvalue problem and $C_1, C_2$ are normalization
constants. These are mixed states, analogous to 
$K_L$ and $K_S$, or $B_H$ and $B_L$ of the mesons.  
We shall call 
the CP--violation from the mixing of states $\delta$,
in analogy to the indirect CP--violation of the $K$--mesons.
Similarly, we call the CP--parameter from the 
vertex corrections $\varepsilon'$.  We know that for
the $K$--Mesons $\varepsilon_K \approx {\cal O}(10^{-3})$ and
$\varepsilon'_K \approx {\cal O}(10^{-7})$.  It will be
interesting to know if there is also a hierarchy
among the various terms contributing to leptogenesis.
\section{Properties of Heavy Majoranas}
We can estimate the probability for collisions of these
states.  A typical interaction is shown in the diagram of Fig. 3.
%DIAGRAM (Fig.3)
\begin{figure}[h]
\centerline{\epsfig{file=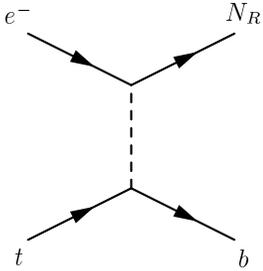,width=3.5cm}}
\caption{A typical scattering}
\end{figure}
Taking the density of states in the early universe to be
$n=\frac{2}{\pi^2} T^3$
and calculating the cross section at an energy $E$ 
equal to the temperature $T$, we obtain
\begin{equation}
n\cdot \sigma\cdot v = \frac{|h_t|^2|h_{\ell}|^2}
                             {8\pi^3}\, T
\end{equation}
with $\sigma$ the cross section, $v$ their relative
velocity and $h_t$, $h_{\ell}$ the couplings of the
Higgses to quarks and leptons, respectively.
At that early time the decay width of the moving leptons 
with mass $M_N$ is
\begin{equation}
\Gamma_N = \frac{|h_{\ell}|^2}{16\pi}\frac{M_N^2}{T}
\end{equation}
leading to 
\begin{equation}
\frac{n\cdot \sigma \cdot v}{\Gamma_N} \sim
 \left(\frac{T}{M_N}\right)^2|h_t|^2.
\end{equation}

Thus, at early stages of the universe with $T\gg M_N$,
when the mixed states are created, they live long enough so that
in one life--time they have many interactions with
their surroundings.  They develop into incoherent
states \cite{Covi,Flanz96}.

When they decay, they produce more leptons than
antileptons.  The excess appears in each one of the decays, 
but does not survive on the large scale of the universe,
because the inverse reaction (recombinations) washes it 
out. The excess survives when the recombinations
cease to take place, as it happens when they decouple.

As the universe expands its temperature and the 
energy of the particles decrease.  At some
time the energy of each particle becomes smaller than
half the mass of the heaviest neutrino.  At that stage 
the heaviest
neutrino decouples.  These neutrinos decay, but cannot
be reproduced because the decay products do not have 
enough energy.  Over the course of time the energy of
each particle becomes smaller than half the mass of
the lightest Majorana neutrino.  As a result the 
neutrinos decay but cannot be reproduced.  The universe
deviates from thermal equilibrium.  Every heavy neutrino
decays and leaves a signature of its presence in the 
excess of the produced leptons.

It is interesting to ask if there is a signal of
this primordial CP--violation which can be observed
at low energies.  

The dynamical creation of the asymmetry appears in
the presence of many other particles of the early
universe.  They are out of thermal equilibrium and
develop according to the Boltzmann equations.  The
surviving asymmetry in the final state depends on the
ratio
\begin{equation}
K=\left( \frac{\langle\Gamma_{\psi_1}\rangle}
    {H(z)}\right)_{T=M_1}
\end{equation}
with $H$ the Hubble constant at temperature $T=M_1$.
Solutions of the Boltzmann equations \cite{Flanz96,Buch}
give the development of the asymmetry as function
of the inverse temperature: $z = M_1/T$.  The development
is shown in figure 4 \cite{Flanz96} for three values
of the parameter $K$.  At a temperature smaller than
$M_1$, i.e. at an epoch which is later than the
time of the $N_1$ decays, the asymmetry reaches constant
asymptotic values $F$, which should not be much
smaller than $F\simeq 10^{-3}$.

Numerical studies have shown that a consistent 
picture emerges provided that
\begin{enumerate}
\item[(i)] the dilution factor in the 
out--of--equilibrium decays is $F \sim 10^{-3}$, and
\item[(ii)] the asymmetry $\varepsilon$ from
individual decays is of order $10^{-4}$ to $10^{-5}$ for
$g_*=100$ degrees of freedom.
\end{enumerate}
The generated lepton asymmetry survives down to
the electroweak phase transition, where a fraction
is converted to a baryon asymmetry.
This happens in terms of topological solutions of field theories 
(sphalerons) and decreases the asymmetry by a factor 
$\frac{28}{79}$.

To sum up, in order to obtain a large lepton and subsequently baryon
asymmetry, three conditions must be satisfied.
\begin{enumerate}
\item[1)]  The Leptogenesis must occur after Inflation, so
that it is not diluted.  This gives the condition
$M_{N_1}<10^{15}$ GeV.
\item[2)]  For the states to be incoherent
\begin{displaymath}
\frac{n\cdot\sigma\cdot v}{\Gamma_N}>P
\end{displaymath}
with $P$ a large number.
For $P=10^3$ this condition gives
\begin{equation}
T>\sqrt{P} M_{N_1}\approx 30 M_{N_1}
\end{equation}(Incoherence Condition)
\item[3)] The dilution factor from thermal development
should not be too large
or too small.  Acceptable values are
$F\sim 10^{-3}$ to $10^{-4}$ for
\begin{displaymath}
K<10^{-3} \quad\quad {\rm to} \quad\quad 10^{-4}\, .
\end{displaymath}
This is known as the out--of--equilibrium condition
and leads to $M_{N_1} > 10^7$ GeV.
\end{enumerate}
\begin{figure}[h]
\epsfig{file=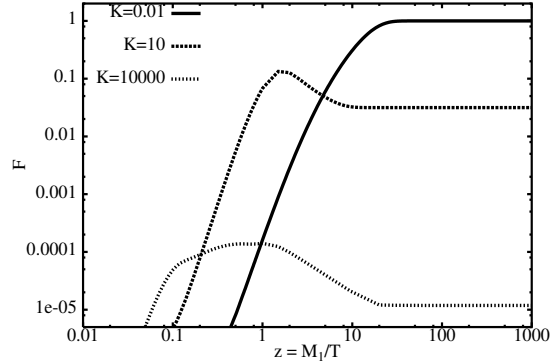,width=7cm,height=5cm}
\caption{Dilution factor $F$ (from the second reference in \cite{Flanz96})}
\end{figure}
%Figure 'Dilution Factor F'
All three conditions can be satisfied in the early
universe with a large range for the masses still being
possible.

The lepton asymmetry created by the above method can
be converted into a Baryon asymmetry at the electroweak
scale.  This phenomenon takes place through topological
solutions of non--abelian gauge theories.  The new
solutions are called sphalerons and create a baryon
excess \cite{Khlepnikov}
\begin{equation}
\frac{n_B}{S}=-\frac{8N_G+4N_H}{22N_G+13N_H}
  \left(\frac{n_L}{S}\right)_{\rm initial}
\end{equation}
with $n_L$ and $n_B$ the excess of leptons and baryons,
respectively.  They are normalized to the entropy $S$
with 
$N_H$ is the number of Higgs doublets, and
$N_G$ the number of generations in the theory.

For typical theories $N_H=1$ and $N_G=3$, leading to
\begin{equation}
\frac{m_B}{S} \approx \frac{1}{3} \frac{n_L}{S} =
 \frac{1}{3} \frac{\delta+\varepsilon'}{g}\, F\, .
\end{equation}
With the numerical values, mentioned in the previous
section, we obtain a Baryon asymmetry consistent
with the one observed in the universe.

\section{Possible Observables}

Leptogenesis will become more interesting and
important when it will be related to other physical 
phenomena. Two interesting questions arise in
this respect:
\begin{enumerate}
\item[1)] Are the CP asymmetries observed in low
energy laboratory experiments related to the 
CP--violation in the early universe?
\item[2)] Are there other macroscopic remnants of the 
cosmological CP--violation, besides the matter asymmetry,
which we can observe?
\end{enumerate}
These are important questions whose consequences are
beginning to emerge.  
In the past year, it was recognized that for specific
models there are relations between the high and low
energy phenomena.
There are already models where a connection has been
established.  

The Majorana mass matrix is symmetric
and can be diagonalized by a unitary matrix $U_R$
\begin{equation}
U_RM_RU_R^+={\rm diag.}\quad(M_1,M_2,M_3).
\end{equation}
On this basis the current states $N'_R$ are related to
the physical states $N_R$ by $N_R=U_R N'_R$.
The transformation also changes the Dirac mass matrix  
\begin{displaymath}
m_D = \tilde{m}_D U_R^+
\end{displaymath}
where $\tilde{m}_D$ is the original Dirac mass matrix
appearing in the Lagrangian. This demonstrates how the
right--handed mixing  matrix enters the Dirac mass matrix
and consequently the lepton asymmetry.

The low energy phenomena, on the other hand, are 
determined by the matrix
\begin{equation}
m_{\nu} = -m_D\,\,\,{\rm diag.}
\left(M_1^{-1},M_2^{-1},M_3^{-1}\right) m_D^T.
\end{equation}
In fact, $m_{\nu}$ is determined by the masses, mixing
angles and phases occurring in low energy phenomena.
In specific models for $\tilde{m}_D$ and $M_R$,
it is possible to invert Eq.\ (12), to obtain 
$m_R$ and consequently the
lepton asymmetry $\varepsilon$ in Eq.\ (3).  
This has been done
in several articles \cite{Ellis}--\cite{Falcone} 
and has been discussed
in a more general framework of $SU(2)\times U(1)$ 
\cite{Branco}.  
Many of these  models consider $SU(2)\times U(1)$ theories
with the see--saw mechanism.

An alternative approach considers the left--right 
symmetric models based on $SU(2)_L\times SU(2)_R$.
The left-- and the right--handed mass matrices are
now related  
\begin{equation}
m_L=f v_L\quad\quad{\rm and}\quad\quad m_R=f v_R
\end{equation}
with $f$ the entries of the mass matrices and
$v_L,\, v_R$ the vacuum expectation values. 

In case
that the low energy phenomena are determined by
$m_L$, the lepton asymmetry is predicted \cite{Joshi}. 
Fig.\ 5 shows the results of an analysis \cite{Joshi} 
where $m_L$ is determined by the observed neutrino
mass differences and mixing angles.  In particular,
the analysis adopted the hierarchical mass scheme
and calculated the asymmetry for the three solar
solutions.  We note that the large--mixing--angle
and the vacuum--oscillation solutions produce
acceptable values for the baryon asymmetry over
an extended region of $\sin^2\theta_3$.

\begin{figure}
\epsfig{file=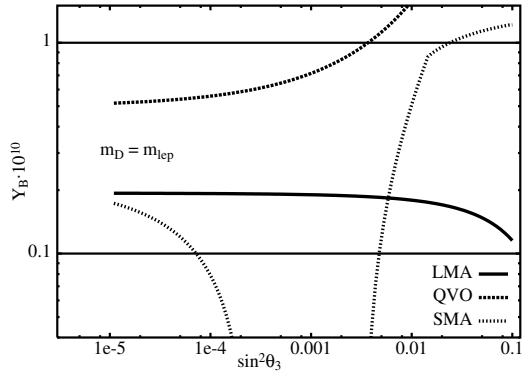, width=7cm, height=5cm}
\caption{Region allowed for the asymmetry}
\end{figure}
%Figure 'Region allowed for the asymmetry'

In the same  model, it is possible to calculate the
lepton mass parameter
\begin{displaymath}
\langle m_{ee}\rangle=\sum_i U_{ei}^2 m_i
\end{displaymath}
entering the neutrinoless double beta decay.  In
special cases large values of 
\begin{displaymath}
\langle m_{ee}\rangle \sim 10^{-3}\quad
 {\rm to} \quad 10^{-2} \rm{ eV}
\end{displaymath} 
are allowed which are close to the bound
established in the  Heidelberg--Moscow
experiment \cite{Klapdor}
$\langle m_{ee}\rangle< 0.35$ eV. 

\section{Summary}

Neutrino physics remains an active field of research
with the development of new topics for investigation.
An open issue is the origin of the unique properties
of neutrinos.  We would like to know the nature
of the neutrinos:  are they Dirac or Majorana
particles?  In case they are Majorana particles,
there is the attractive possibility that they
generated a lepton--asymmetry in the early universe, 
which was later converted to a baryon asymmetry.

Several articles have been published during the past
year, relating the masses and mixings observed in the
oscillation experiments to the high energy phenomena 
that took place in the early universe.  This is a 
welcome development, because several apparently remote
phenomena seem to be related to each other.

\end{document}